\documentclass[aps,twocolumn]{revtex4}
\usepackage[dvips]{graphics,graphicx}
\usepackage{amsmath}
\begin{document}

\title{Open Fermi-Hubbard model: Landauer's {\em vs.} master equation approaches}
\author{Andrey R. Kolovsky$^{1,2}$}
\affiliation{$^1$Kirensky Institute of Physics, 660036 Krasnoyarsk, Russia}
\affiliation{$^2$Siberian Federal University, 660041 Krasnoyarsk, Russia}
\date{\today}

\begin{abstract}
We introduce a simple model for the quantum transport of Fermi particles between two contacts connected by a lead. It  generalizes  the Landauer formalizm by explicitly taking into account the relaxation processes in the contacts.  We calculate the contact resistance and non-equilibrium  quasi-momentum distribution of the carriers in the lead and show that they strongly depend on the rate of relaxation processes.   
\end{abstract}
\maketitle

\section{Introduction}
\label{sec1}
Recently much attention has been paid to dynamics and non-equilibrium states of open many-body systems \cite{Bran12,Brud12,Ivan13,Niet14,Pros14,Simp14,Labo15,Kord15b,Znid15,Buca17,108,Lebr18,112,114,Xhan20,116}. 
Here the term `open' means that the system of interest is coupled to a bath and, hence, generally neither the system  energy nor the number of particles in the system are conserved.  Typical examples of open many-body systems are the open Fermi-Hubbard and Bose-Hubbard models \cite{Ivan13,Pros14,Kord15b,Buca17,112,114,116} which are supposed to describe the current of fermionic or bosonic particles between two reservoirs (the contacts) connected by a one-dimensional lattice (the lead).  The mathematical framework of the models is the master equation for the reduced density matrix of the carriers in the lattice with two relaxation terms acting on the first and the last  sites of the lattice. Remarkably,  these models can be tackled analytically or semi-analytically, leading to a number of important conclusions. In particular, it was shown in the recent work \cite{116} that in the case of Bose particles the inter-particle interactions result in a change of the ballistic transport regime, where the current across the lattice is independent of the lattice length $L$, to the diffusive transport, where the current is inverse proportional  to $L$.

Although the open Fermi- and Bose-Hubbard models are important in the field of quantum transport, they have a limited applicability because they rely on the Markovian master equation which is only justified for high-temperature reservoirs \cite{108,114,117}. The case of low-temperature particle reservoirs, which is of particular interest in solid-state physics, remains a challenge. A popular approach to a non-Markovian bath is the stochastic Schr\"odinger equation with the correlated noise \cite{Dios98,Zhao12,Chen13,Sues15}. As shown in Ref.~\cite{Sues15}, this leads to an infinite set of the coupled Lindblad-like master equations which should be truncated to a finite set to ensure a given accuracy. Unfortunately,  the application of this method to the open Hubbard chains looks unfeasible for the moment.  In the present work we explore a different approach which allows us to stay within the Markovian approximation in spite of the fact that the reduced density matrix of the carriers in the chain does not obey a Markovian master equation.  The idea is to use the hierarchical  reservoirs where the contacts are both part of the system and the larger reservoirs.

The structure of the paper is the following. In the next section we introduce a simple model where the Hubbard chain is dressed by the contacts. In Sec.~\ref{sec3}  we analyze the current of the Fermi particle across the chain as the function of the model parameters and calculate non-equilibrium distributions of the carriers over the Bloch states for the carriers in the contacts and the lead. The relation to the Landauer equation is discussed in Sec.~\ref{sec4}.  
Finally, the concluding Sec.~\ref{sec6} summarizes the obtained results and indicates some prospects of the further research. 

\section{Model}
\label{sec2}
\begin{figure}[b]
\includegraphics[width=8.5cm,clip]{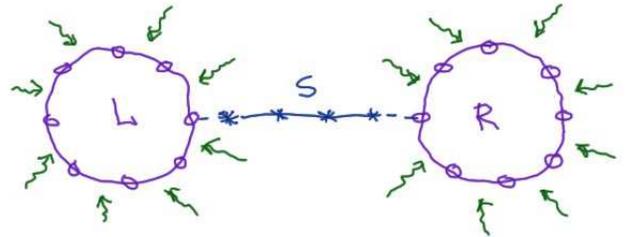}
\caption{Schematic presentation of the model. Wavy arrows indicate the particle exchange between the contacts and reservoirs.}
\label{fig1}
\end{figure}

Let us consider two contacts connected by the Hubbard chain (see Fig.~\ref{fig1}),
\begin{equation}
\label{a0}
\widehat{H}=\widehat{H}_{\rm L} + \widehat{H}_{\rm R} + \widehat{H}_s + \widehat{H}_{\epsilon}^{({\rm L})} + \widehat{H}_{\epsilon}^{({\rm R})} \;.
\end{equation}
In Eq.~(\ref{a0}) $\widehat{H}_{\rm L} $ and $\widehat{H}_{\rm R}$ are Hamiltonians of  the left and right contacts,   $\widehat{H}_s$ is the Hamiltonian of the carriers in the chain, and $\widehat{H}_{\epsilon}^{(j)}$, where $j={\rm L},{\rm R}$, are the coupling Hamiltonians. 

The Hamiltonians of the contacts read
\begin{equation}
\label{a1}
\widehat{H}_j= \sum_k E_k \hat{b}_k^\dagger \hat{b}_k  \;, \quad  E_k=-J\cos\left(\frac{2\pi k}{M}\right) \;,
\end{equation}
where  $\hat{b}_k^\dagger$ and $\hat{b}_k$ are the creation and annihilation operators which create or annihilate a particle in the Bloch state with the quasimomentum $\kappa=k/M$. Notice that these operators, as well as the Hamiltonian parameters, also carry the index $j$ which we omit here not to  overburden the equation. The contacts are assumed to be a part of the larger particle reservoirs which enforce the relaxation of the reduced density matrices $\hat{R}^{(j)}(t)$ of the isolated contacts into the equilibrium state given by the Fermi-Dirac distribution for the fermionic carriers and Bose-Einstein distribution for the bosonic carriers,
\begin{equation}
\label{a2}
 n_k={\rm Tr}[\hat{b}_k^\dagger \hat{b}_k \hat{R}(t=\infty)]=\frac{1}{e^{\beta(E_k-\mu)} \mp 1}      \;.
\end{equation}
To be certain, from now on we shall consider the spinless fermions and zero reservoir temperature. Then the explicit form of the Lindblad relaxation operator is
\begin{equation}
\label{a3}
{\cal L}(\hat{R})=-\frac{\gamma}{2}\sum_{|k|<k_F} \left( \hat{b}_k \hat{b}_k^\dagger \hat{R} 
-\hat{b}_k^\dagger \hat{R}  \hat{b}_k  +  \hat{R} \hat{b}_k^\dagger \hat{b}_k \right)  \;,
\end{equation}
if $|k|<k_F$ and
\begin{equation}
\label{a4}
{\cal L}(\hat{R})=-\frac{\gamma}{2}\sum_{|k|>k_F} \left( \hat{b}_k^\dagger \hat{b}_k \hat{R}
-\hat{b}_k \hat{R}  \hat{b}_k^\dagger  +  \hat{R} \hat{b}_k \hat{b}_k^\dagger \right)  \;,
\end{equation}
if $|k|>k_F$, where $k_F$ is determined by the Fermi energy  of the corresponding reservoir through the relation $E_F=-J\cos(2\pi k_F/M)$ and $\gamma$ is the relaxation constant. 

For fermions in the chain we elect to work in the Wannier basis. Then the chain Hamiltonian is given by the tight-binding model for the spinless fermions,
\begin{equation}
\label{a5}
\widehat{H}_s=-\frac{J}{2}\left( \sum_{l=1}^{L-1} \hat{c}_{l+1}^\dagger \hat{c}_l + h.c. \right)   \;,
\end{equation}
where operator  $\hat{c}_l^\dagger$ ($\hat{c}_l$) creates (annihilates) a fermion in the $l$-th site of the chain.  To simplify the analysis we assume that the hopping constant $J$ in Eq.~(\ref{a5}) coincides with that in Eq.~(\ref{a1}).  This allows us to use $J$ as the energy measurement unit.  

Finally, the coupling operator between the left contact and the chain is
\begin{equation}
\label{a6}
\widehat{H}_\epsilon^{({\rm L})}=\frac{\epsilon}{\sqrt{M}} \left( \hat{c}_{1}^\dagger \sum_{k=1}^{M} \hat{b}_k e^{i\frac{2\pi m}{M} k} + h.c. \right)   \;,
\end{equation}
and the coupling operator between the chain and the right contact has similar form where the operator  $\hat{c}_{1}^\dagger$ is substituted by the operator $\hat{c}_{L}^\dagger$.

The evolution of the system (\ref{a0}) is assumed to obey the Markovian master equation
\begin{equation}
\label{b1}
\frac{{\rm d} \hat{R}}{{\rm d} t}=-i[\widehat{H},\hat{R}] +  {\cal L}_{\rm L}(\hat{R})  + {\cal L}_{\rm R}(\hat{R}) \;,
\end{equation}
where $\hat{R}=\hat{R}(t)$ now denotes the total density matrix of the composed system `contacts$+$chain'. In the considered case of the spinless fermions the size  of this matrix is obviously given by the equation,
\begin{equation}
\label{b2}
{\cal N}=\sum_{n=0}^{N} \frac{N!}{n!(N-n)!}=2^N \;,
\end{equation}
where the parameter $N$ is the total number of the single-particle states, $N=M_{\rm L} +L+M_{\rm R}$. The density matrix $\hat{R}$ carries full information about the system which we actually do not need for our purposes. Indeed, to predict the current between the contacts it suffices to know the single particle density matrix (SPDM) of the size $N\times N$ which is defined according to the equation 
\begin{equation}
\label{b3}
\rho_{k,l}^{(i,j)}(t)={\rm Tr}[\hat{d}^{\dagger (i)}_k \hat{d}^{(j)}_l  R(t)]  \;.
\end{equation}
(Here we use the common notation for the creation and annihilation operators appearing in the problem, where the superindexes $i$ and $j$ now take one of the three meaning  -- ${\rm L}$ for the left contact, $s$ for the chain, and ${\rm R}$ for the right contact.) Our particular interest is the block $\rho_{l,m}^{(s,s)}$ which is the SPDM of the carriers in the chain.  Knowing this block one finds the current as
\begin{equation}
\label{b4}
 j(t)= {\rm Tr}[\hat{j} \hat{\rho}^{(s,s)}(t) ] \;, 
\end{equation}
where $\hat{j}$ is the current operator, $j_{l,m}=J(\delta_{l,m+1} - \delta_{l+1,m})/2i$. Alternatively, one finds the current by using the equation
\begin{equation}
\label{b5}
 j(t)= 2\sum_{k>0} J\sin\left( \frac{2\pi k}{M}\right) f(k,t) \;,
 \end{equation} 
\begin{equation}
\label{b6} 
f(k,t)=\tilde{\rho}^{(s,s)}_{k,k}(t)-\tilde{\rho}^{(s,s)}_{-k,-k}(t)  \;, 
\end{equation}
where $\tilde{\rho}^{(s,s)}$ is the matrix  $\hat{\rho}^{(s,s)}$ in the momentum representation, i.e., the Fourier transform of  $\hat{\rho}^{(s,s)}$.

Next we use the fact that the master equation ({\ref{b1}) has a quadratic form with respect to creation and annihilation operators. In this case one can obtain a closed set of equations for the SPDM elements. Substituting Eq.~(\ref{b3}) into Eq.~(\ref{b1}) we get
\begin{equation}
\label{c1}
\frac{{\rm d} \rho^{(i,j)}_{k,l}}{{\rm d} t}=-i[\widehat{H},\rho]^{(i,j)}_{k,l} - \gamma B^{(i,j)}\rho^{(i,j)}_{k,l} +\gamma A^{(i,j)}_{k,l} \;,
\end{equation}
where $B^{({\rm L},{\rm L})}=B^{({\rm R},{\rm R})}=B^{({\rm L},{\rm R})}=B^{({\rm R},{\rm L})} =1$,  $B^{(s,s)}=0$,   $B^{(s,{\rm L})}=B^{({\rm L},s)}=B^{(s,{\rm R})}=B^{({\rm R},s)}=0.5$,  and $A^{(i,j)}_{k,l}=0$ except the elements $A^{(j,j)}_{k,k}$ which are equal to unity for $|k|<k_F^{(j)}$ of the respective contact. It is easy to see from Eq.~(\ref{c1}) that for vanishing coupling constant $\epsilon$ the density matrices of the contacts relax to the diagonal matrices with the diagonal elements obeying the Fermi-Dirac distribution (\ref{a2}). However, if $\epsilon\ne0$ and $k_F^{({\rm R})} \ne   k_F^{({\rm L})}$ the system relaxes to a non-equilibrium state with the stationary current $\bar{j}$ flowing between the contacts. In what follows we analyze this non-equilibrium state in more detail.

\section{Numerical results}
\label{sec3}
We solve Eq.~(\ref{c1}) numerically for different system size and different parameter values. The panels (a) and (b) in Fig.~\ref{fig2} illustrate relaxation of the system to the steady state for $M_{\rm L}=M_{\rm R}=L=60$, $J=1$, $\epsilon=0.5$, $E_F^{({\rm L})}=0.3$, $E_F^{({\rm R})}=-0.3$, and  $\gamma=0.05$.  The panel (a) shows population dynamics of the lattice sites in the situation where initially there were no particles in the system. It is seen that the site occupations $n_l(t)$ slowly approach the value 0.5. Unlike this slow process, the mean current  $j(t)$ rapidly reaches the stationary value $\bar{j}/L\approx0.06$. Thus, there are two  characteristic relaxation times in the system,  $\tau_1$ and $\tau_2\gg \tau_1$, which scale differently with the chain length $L$.  The system reaches its true steady state for $t>\tau_2$, which for the chosen $L$ and the initial condition is about $10^4$ tunneling periods.
\begin{figure}[t]
\hspace*{-1cm}
\includegraphics[width=10.5cm,clip]{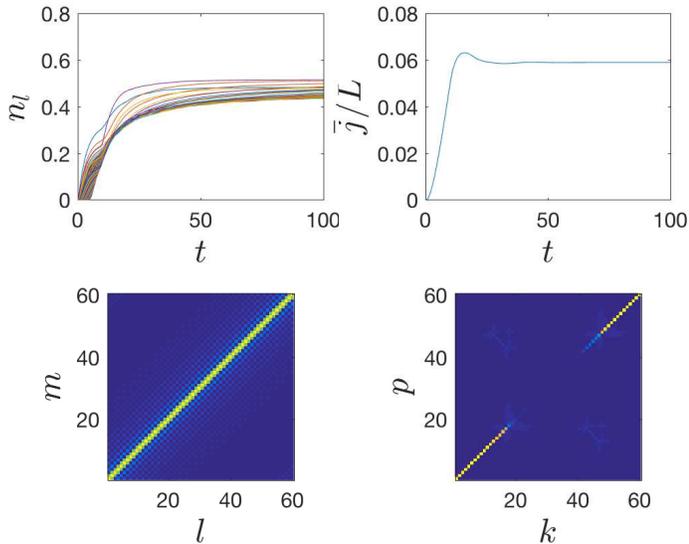}
\caption{Upper row: Populations of the chain sites (left) and the mean current normalized to the chain length (right) as the functions of time which is measured in the units of the tunneling period. Lower row:  Single-particle density matrix of the carriers in the chain at $t=10^4$ in the coordinate (left) and momentum (right) representation. Parameters are $M_{\rm L}=M_{\rm R}=L=60$, $J=1$, $\epsilon=0.5$,  $E_F^{({\rm L})}=0.3$, $E_F^{({\rm R})}=-0.3$, and $\gamma=0.05$. Initial condition corresponds to the empty system.}
\label{fig2}
\end{figure} 
\begin{figure}
\includegraphics[width=8.5cm,clip]{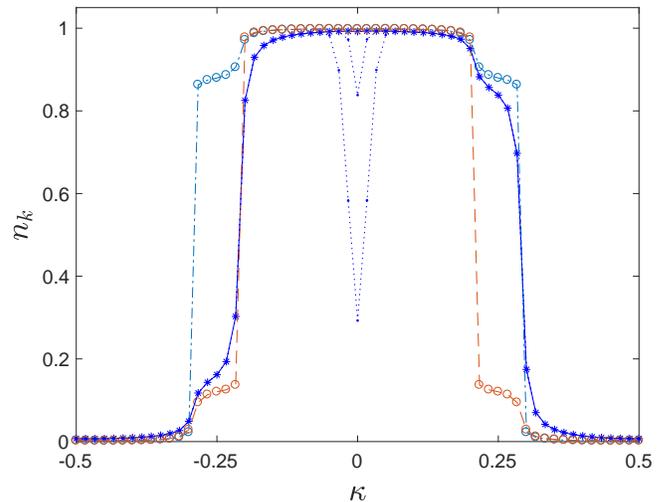}
\caption{Stationary momentum distributions (i.e., occupation numbers of the Bloch states) for the carriers in the left contact, dash-dotted line, in the right contact, dashed line, and in the chain, solid line. Additional dotted lines show the momentum distribution of the carriers in the chain at $t=250$ and $t=2500$.}
\label{fig3}
\end{figure} 

Next we discuss the stationary SPDM of the carriers in the chain. The lower panels in Fig.~\ref{fig2} show the matrix $\hat{\rho}^{(s,s)}(t=10^4)$  in the coordinate and momentum representations, respectively.  It is seen that the stationary SPDM is approximately diagonal in the momentum representation, where we plot the values of the diagonal elements in Fig.~\ref{fig3} by asterisks connected by the solid line. Additionally, the dash-dotted and dashed lines in Fig.~\ref{fig3} show occupation numbers of the contact Bloch states.  It is seen that the Fermi-Dirac distributions of the isolated contacts are slightly perturbed by the lead. On the contrary, the momentum distribution of the carriers in the chain strongly deviates from the equilibrium Fermi-Dirac distribution. Namely, it is asymmetric with  respect to the reflection $\kappa \rightarrow -\kappa$.  Due to this asymmetry we have non-zero net current which can be calculated by using Eqs.~(\ref{b5}-\ref{b6}). It is also an appropriate place here to comment on the relaxation time $\tau_2$.  The transient system dynamics for $\tau_1<t<\tau_2$ is reflected in the momentum distribution as a deep at $\kappa=0$ (see dotted lines in Fig.~\ref{fig3}) which disappears only for $t>\tau_2$. However, since this deep is symmetric with respect to the reflection, it affects neither the function $f(\kappa)$ nor the value of the current as soon as $t>\tau_1$.
\begin{figure}
\includegraphics[width=8.5cm,clip]{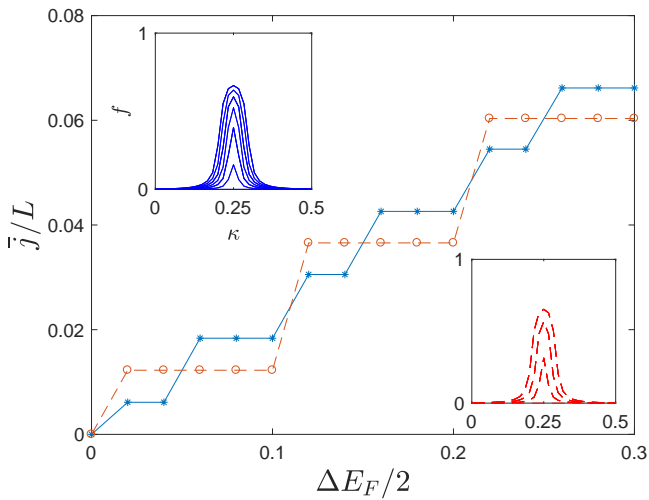}
\caption{Stationary current as the function of the chemical potential difference. The chain length is $L=60$, the contact size are $M_{\rm L}=M_{\rm R}=60$, dashed line, and $M_{\rm L}=M_{\rm R}=120$, solid line. The other parameters are $\epsilon=0.5$,  $E_F^{({\rm R})}=\Delta E_F/2$,  $E_F^{({\rm L})}=-\Delta E_F/2$., and $\gamma=0.125$. The inserts show the function $f(\kappa)$ (only positive part is shown) for every plateau.}
\label{fig4}
\end{figure} 

Finally we analyze the stationary current as the function the system parameters. To be certain we shall assume $E_F^{({\rm R})}=\Delta E_F/2$, and $E_F^{({\rm L})}=-\Delta E_F/2$.  The dashed line in the main panel in Fig.~\ref{fig4} shows the stationary current as the function of $\Delta E_F$ for the system size $M_{\rm L}=M_{\rm R}=L=60$. The observed step-like dependence is due to finite size of the contacts. Indeed, increasing the number of states in the contacts two times we double the number of steps. Thus, in the limit $M_{\rm L},M_{\rm R} \rightarrow \infty$ we get a smooth dependence,
\begin{equation}
\label{c2}
\frac{\bar{j}}{L}\approx G(\epsilon,\gamma) \Delta E_F  \;,
\end{equation}
where the conductance $G=G(\epsilon,\gamma)$, also known as the inverse contact resistance, is some function of the relaxation constant $\gamma$ and the coupling constant $\epsilon$. The dependence (\ref{c2}) is exemplified in Fig.~\ref{fig5}. The left panel in Fig.~\ref{fig5} shows the stationary current as the function of  $\epsilon$  for three different values of the chemical potential difference $\Delta E_F$, where we set the relaxation constant $\gamma=0.1$. The right panel shows the stationary current as the function of $\gamma$ where we set the coupling constant $\epsilon=0.5$.  Additionally, in Fig.~\ref{fig6} we depict the function $f(\kappa)$ which sheds more light on  the observed non-trivial dependence of the current on the relaxation constant $\gamma$.  This dependence is also discussed in Appendix  which contains some analytical results for the stationary SPDM.

\begin{figure}
\includegraphics[width=8.5cm,clip]{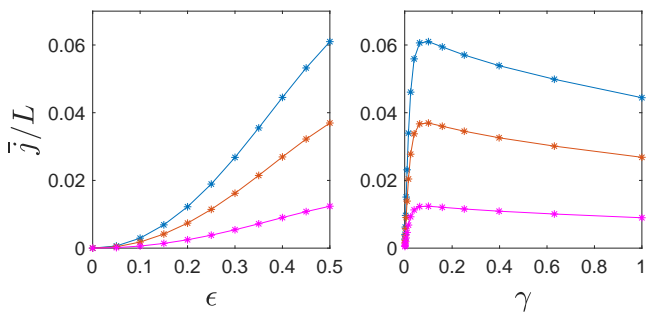}
\caption{Left panel: Stationary current as the function of the coupling constant $\epsilon$ for $\gamma=0.1$ and three values of the parameter $\Delta E_F/2=0.1,0.2,0.3$, from bottom to top. Right panel: Stationary current as the function of the relaxation constant $\gamma$ for $\epsilon=0.5$ and three values of the parameter $\Delta E_F/2=0.1,0.2,0.3$, from bottom to top.}
\label{fig5}
\end{figure} 
\begin{figure}
\includegraphics[width=8.5cm,clip]{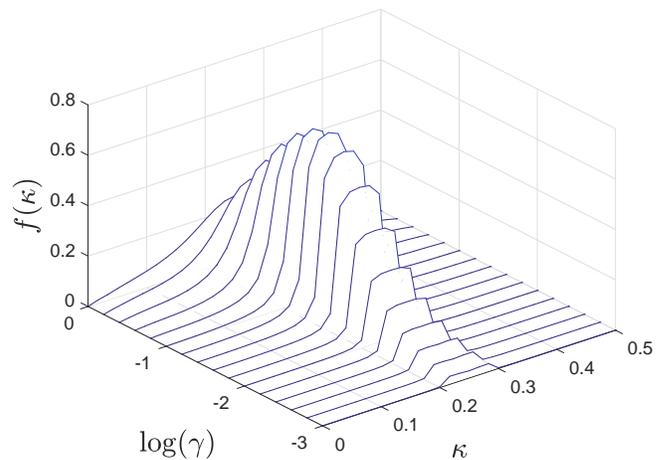}
\caption{The function $f(\kappa)$ for the parameters of Fig.~\ref{fig5}(b) and $\Delta E_F/2=0.3$. Only positive part $\kappa\ge0$ is shown.}
\label{fig6}
\end{figure} 
\begin{figure}
\includegraphics[width=8.5cm,clip]{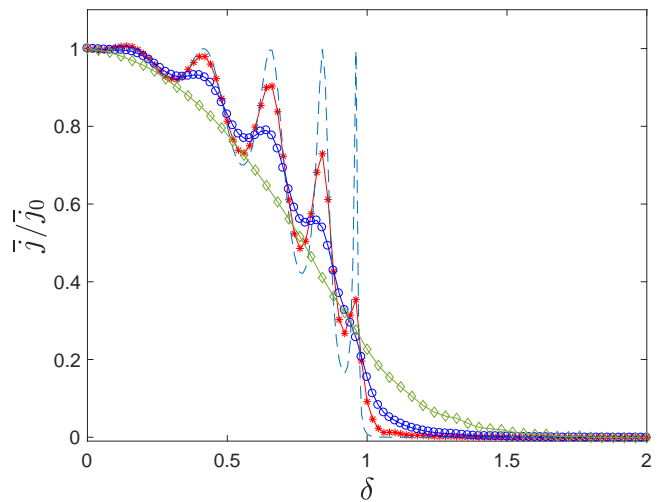}
\caption{The  stationary current across the chain with the square potential barrier of the 11 sites width and the hight $\delta$.  Symbols show the ratio  $\bar{j}/\bar{j}_{0}$ as the function of $\delta$ for $\gamma=0.01$ (asterisks), $\gamma=0.1$ (circles), and $\gamma=0.5$ (diamonds).  The other system parameters are $\epsilon=0.5$ and $\Delta E_F/2=0.1$.  The dashed line is the transmission coefficient of the barrier at $E=E_F=0$.}
\label{fig7}
\end{figure}

\section{Imperfect chain}
\label{sec4}
Till now we analyzed the case of the perfect lead. It is interesting to consider an imperfect lead where, according to the Landauer arguments, the stationary current should obey the equation 
\begin{equation}
\label{c3}
\bar{j} = \bar{j}_{0} |t(E_F)|^2 \;,
\end{equation}
where  $t(E_F)$ is the transmission amplitude for the imperfect chain at the Fermi energy and $\bar{j}_{0}$ is the stationary current in the perfect chain. However, Eq.~(\ref{c3}) neglects the decoherence effect of reservoirs by approximating  the quantum state of the carriers by the plane wave with the quasimomentum $k_F$.  To discuss the validity of Eq.~(\ref{c3})   we simulate  the system dynamics for the chain with the square potential barrier.  The dashed line in Fig.~\ref{fig7} shows the transmission coefficient of the barrier as the function of its hight $\delta$.  A number of transmission peaks and deeps due to the phenomenon of the resonant above-barrier reflection are clearly seen. The solid lines in Fig.~\ref{fig7} are the ratio $\bar{j}/\bar{j}_0$ for the three different values of the relaxation constant $\gamma$. It is seen that the transmission peaks are washed out if $\gamma$ is increased.  Thus, Eq.~(\ref{c3}) can be valid only in the limit $\gamma\rightarrow 0$ where, according to the results of Fig.~\ref{fig5}(b), the stationary current vanishes.

\section{Conclusion}
\label{sec6}
We introduced a simple model for the transport of Fermi particles between two contacts with different chemical potentials. The numerical analysis of the model shows that its properties fit well the Landauer approach for the electron transport in the mesoscopic devices \cite{Datt95}. In particular, all relaxation processes in the system take place at the contacts. The main difference with the Landauer approach is that we describe these processes explicitly by using the formalizm of the master equation. This allows us to relax the assumption about the `reflectionless' contacts and calculate the non-equilibrium distribution of the carriers over the Bloch states for arbitrary value of the relaxation constant $\gamma$  and the coupling constant $\epsilon$ -- the parameters which are absent in the  standard Landauer theory.  Since the constant $\gamma$ also determines the rate of decoherence in the system, one can address within the framework of the introduced model a number of other questions like, for example, the decoherence effect of reservoirs on the Anderson localization in a disordered chain. 

The other prospect of  the research is the non-Markovian master equation. For the considered problem one obtains such equation by considering the low-temperature limit and eliminating the contacts. The analysis of this non-Markovian master equation (including various approximations) is of considerable academic interest.

The author acknowledges discussions with D. N. Maksimov and financial support of Russian Science Foundation (RU) through Grant No.19-12-00167.


\newpage
\section{Appendix}
This section contains some analytical results for the stationary SPDM which was analysed numerically in Sec.~\ref{sec3}. For the sake of completeness  we also include results for infinite reservoir temperature.

\subsection{The hight-temperature limit} 
We begin with the case of infinite temperature where the Bloch states of the contacts are equally occupied. The mean occupation numbers $\bar{n}_{\rm L}$ and   $\bar{n}_{\rm R}$ are obviously determined by the reservoir chemical potentials and, as before, we assume  $\mu_{\rm L}>\mu_{\rm R}$ that implies $\bar{n}_{\rm L}>\bar{n}_{\rm R}$. Assuming additionally $\gamma\gg\epsilon^2$ one justifies the Born-Markov approximation which allows us to obtain the master equation for the reduced density matrix of the carriers in the chain,
\begin{equation}
\label{d1}
\frac{d \hat{\rho}}{dt}=-i[\widehat{H}_s,\hat{\rho}] + {\cal L}_{\rm L}(\hat{\rho}) + {\cal L}_{\rm R}(\hat{\rho}) \;, 
\end{equation}
where $\hat{\rho}(t)\equiv\hat{\rho}^{(s,s)}(t)$ and the Hamiltonian $\widehat{H}_s$ is given in Eq.~(\ref{a5}). The explicit form of the Lindblad operator ${\cal L}_{\rm L}(\hat{\rho})$ is
\begin{eqnarray}
\label{d2}
{\cal L}_{\rm L}(\hat{\rho})=-\frac{\tilde{\gamma}}{2} \left[ (1-\bar{n}_{\rm L})
(\hat{c}_1^\dagger \hat{c}_1\hat{\rho}-2\hat{c}_1\hat{\rho}\hat{c}_1^\dagger + \hat{\rho}\hat{c}_1^\dagger\hat{c}_1) \right. \\
\nonumber
\left. \bar{n}_{\rm L}(\hat{c}_1 \hat{c}_1^\dagger \hat{\rho}-2\hat{c}_1^\dagger\hat{\rho}\hat{c}_1 
+ \hat{\rho}\hat{c}_1\hat{c}_1^\dagger)  \right] \;,
\end{eqnarray}
%
\begin{figure}[b]
\includegraphics[width=8.5cm,clip]{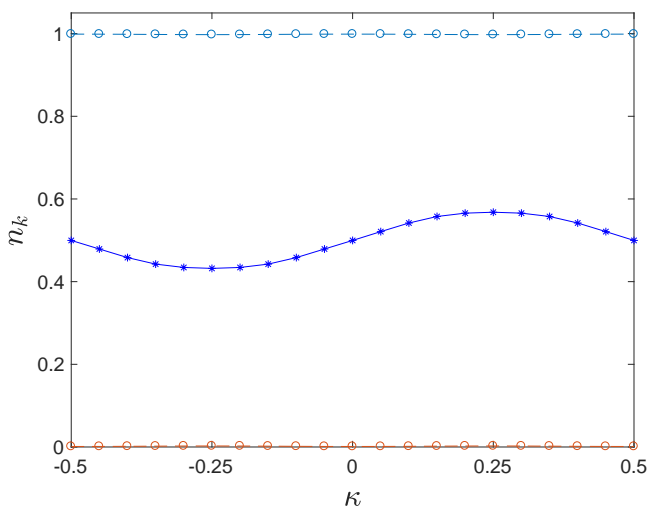}
\caption{Stationary momentum distributions  for $\beta=0$. The system parameters are $M_{\rm L}=M_{\rm R}=L=20$,  $\bar{n}_{\rm L}=1$, $\bar{n}_{\rm R}=0$,  $\epsilon=0.2$, and $\gamma=1.0$.}
\label{fig8}
\end{figure} 
where the effective relaxation constant $\tilde{\gamma}\sim \epsilon^2$. The operator ${\cal L}_{\rm R}(\hat{\rho})$ has the similar form with the creation and annihilation  operators $\hat{c}^\dagger_1$ and $\hat{c}_1$ substituted by the operators  $\hat{c}^\dagger_L$ and $\hat{c}_L$. The master equation (\ref{d1}) with the specified relaxation terms can be solved analytically \cite{114} that gives
\begin{equation}
\label{d3}
 n_k^{(s)}=A+ 2B \sin\left(\frac{2\pi k}{L}\right) \;,   
 \end{equation}
where 
\begin{equation}
\label{d4}  
 A=\frac{ \bar{n}_{\rm L} + \bar{n}_{\rm R}}{2} \;,\quad
 B=\frac{\tilde{\gamma}J}{\tilde{\gamma}^2+J^2}  \frac{ \bar{n}_{\rm L} - \bar{n}_{\rm R}}{2}  \;. 
\end{equation}
The analytical results (\ref{d3}) is verified  in Fig.~\ref{fig8} which shows the occupation numbers $n_k^{(j)}$, $j={\rm L}, s, {\rm R}$, obtained by the straightforward  numerical solution of the master equation (\ref{c1}). The sine dependence of $n_k^{(s)}$ on the quasimomentum is clearly observed. The value of the effective relaxation constant extracted from the depicted numerical data is $\tilde{\gamma}=0.07$.

\subsection{The low-temperature limit}
We come back to the case of zero reservoir temperature. Let us first discuss the limit $\gamma,\epsilon\rightarrow 0$. In this limit the occupation numbers of the Bloch states are  
\begin{equation}
\label{d5}
n_k^{({\rm L},{\rm R})}=\left\{
\begin{array}{ccc}
1&,&|k|<k_F^{({\rm L},{\rm R})}\\
0&,&|k|>k_F^{({\rm L},{\rm R})}
\end{array} \right. \;,
\end{equation}
\begin{displaymath}
n_k^{(s)} = \left(n_k^{({\rm L})} + n_k^{({\rm R})} \right)/2  \;,
\end{displaymath}
and the relaxation terms in the master equation (\ref{c1}) can be approximated by the simpler relaxation term
\begin{equation}
\label{d6}
{\cal L}(\hat{\rho})=-\gamma(\hat{\rho}-\hat{\rho}_0) \;,
\end{equation}
where $\hat{\rho}_0$ is diagonal in the Bloch basis with the elements given by Eq.~(\ref{d5}). We mention that this approximation can be justified only for asymptotically large times where $\hat{\rho}(t)$ is close to its stationary value. 
\begin{figure}[t]
\includegraphics[width=8.5cm,clip]{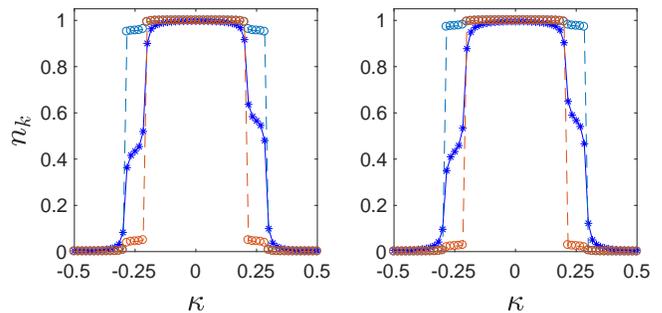}
\caption{Stationary momentum distributions calculated according to Eq.~(\ref{d7}), left, as compared to the exact result, right. The system parameters are  $M_{\rm L}=M_{\rm R}=L=60$,  $\Delta E_F=0.3$, $\epsilon=0.2$, and $\gamma=0.05$. }
\label{fig9}
\end{figure}

Now we set $\epsilon$ to a finite value.  Then, using Eq.~(\ref{d6}), the stationary solution of the master equation reads
\begin{equation}
\label{d7}
\hat{\rho}=\sum_{n,m}  \frac{\gamma\langle \psi_n| \hat{\rho}_0 |\psi_m\rangle}{\gamma+i({\cal E}_m-{\cal E}_n)}
 |\psi_n\rangle\langle \psi_m | \;,
\end{equation}
where $ |\psi_n\rangle$ and ${\cal E}_n$ are eigenenergies and eigenstates of the Hamiltonian (\ref{a0}).  We found that for $\epsilon\ll J$  Eq.~(\ref{d7}) provides a good approximation to the exact stationary SPDM, see Fig.~\ref{fig9}.  

Concluding this subsection we would like to notice correlations between the chain and the contacts which are absent in the case of infinite reservoir temperature. These correlations are exemplified  in  Fig.~\ref{fig10} which shows the stationary SPDM in the original basis (i. e., the Wannier basis for the chain and the Bloch basis for the contacts) as a color map.  In this figure the central block of the size $60\times60$ is the reduced density matrix of the carriers in the chain and the left-lower and right-upper blocks are the reduced density matrices of the carriers in the left and right contacts, respectively.  Besides `contacts-chain' correlations there also are correlations between $+k$ and $-k$ contact Bloch states for $k_F^{({\rm L})} \le |k| \le  k_F^{({\rm R})}$. 
\begin{figure}
\includegraphics[width=8.5cm,clip]{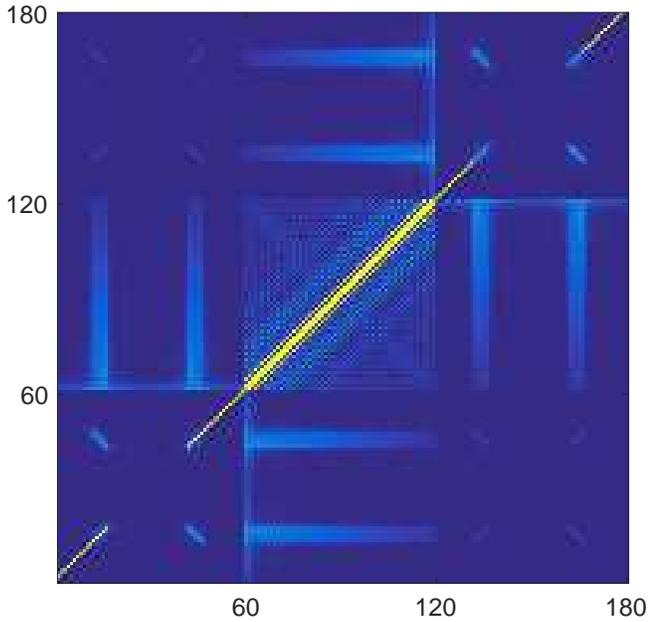}
\caption{The stationary SPDM in the original representation 
Shown are the absolute values of the matrix elements where the upper limit of the color axis is set to 0.1 to highlight the correlations.}
\label{fig10}
\end{figure}

\end{document}